\documentclass[a4paper,aps,prb,twocolumn,floatfix,showpacs,superscriptaddress,footnoteinbib]{revtex4-1}
\usepackage{graphics}
\usepackage{epsfig}
\usepackage{times}
\usepackage{xcolor}   
\usepackage{amsfonts}
\usepackage{amsmath}
\usepackage{amssymb}
\usepackage{amsthm}
\usepackage{hyperref} 
\usepackage{wasysym}
\usepackage{bm}

\begin{document}
\title{Spin Berry phase in a helical edge state: $S_z$ nonconservation and transport signatures}

\author{Vivekananda Adak}
\affiliation{Department of Physical Sciences, IISER Kolkata, Mohanpur, West Bengal 741246, India.}
\author{Krishanu Roychowdhury}
\affiliation{Department of Physics, Stockholm University, SE-106 91 Stockholm, Sweden.}
\affiliation{Laboratory of Atomic and Solid State Physics, Cornell University, Ithaca, New York 14853, USA.}
\author{Sourin Das}
\affiliation{Department of Physical Sciences, IISER Kolkata, Mohanpur, West Bengal 741246, India.}
\email{vivekanandaaadak@gmail.com, krishanu.1987@gmail.com, sdas@physics.du.ac.in}

\begin{abstract}
Topological protection of edge state in quantum spin Hall systems relies only on time-reversal symmetry. Hence, $S_z$ conservation on the edge can be relaxed which can have an interferometric manifestation in terms of spin Berry phase. Primarily it could lead to the generation of spin Berry phase arising from a closed loop dynamics of electrons. Our work provides a minimal framework to generate and detect these effects by employing both spin-unpolarized and spin-polarized leads. We show that spin-polarized leads could lead to resonances or anti-resonances in the two-terminal conductance of the interferometer. We further show that the positions of these anti-resonances (as a function of energy of the incident electron) get shifted owing to the presence of spin Berry phase. Finally, we present simulations of a device setup using KWANTpackage which put our theoretical predictions on firm footing. 
\end{abstract}
\maketitle


\section{Introduction}

Birth of topological insulators~\cite{kane2005quantum, moore2007topological, roy2009topological, joel2010birth, xiao2011insulators} has marked a new realm in the field of condensed matter research and nucleated a number of experimental activities~\cite{hasan2010insulators, PhysRevLett.107.086803, PhysRevB.83.165304} in quest for materials relevant for exploring the topological aspects of such systems in the past decades. Endowed with an exotic surface physics, these materials~\cite{ando2013topological} can be described in terms of simple band Hamiltonians with spin-orbit (SO) couplings which respect time-reversal symmetry. The surface states owe their existence to the nontrivial topology of the bulk as an implication of bulk-boundary correspondence~\cite{Essin2011bulk}. In two-dimensions, a simplistic description of topological insulators can be captured in the so-called Bernevig-Hughes-Zhang (BHZ) model of HgTe quantum well~\cite{Bernevig1757}. Exceeding a critical value of the well width, an inversion between the bands near the Fermi surface drives the system into a topological insulator state with localized edge modes on the boundary. These edge modes have conserved spin quantum number $S_z$ locked with their momentum viz. if $\uparrow$-spins ($S_z=+1$) flow along $+k$, called right movers, $\downarrow$-spins ($S_z=-1$) would flow along $-k$, called left movers, ensued from time-reversal symmetry -- a phenomenon known as {\it quantum spin Hall} (QSH) effect~\cite{Murakami1348,vzutic2004spintronics,kane2005quantum, sinova2013,Bernevig1757,Konig766,Roth294}. 

The edge state in the BHZ model has linear dispersion around the $\Gamma$ point with conserved helicity ($\propto {\bf S}\cdot{\bf k}$), hence, known as helical edge state (HES)~\cite{PhysRevLett.96.106401}. The spin quantization axis of the HES is aligned along the SO field operative perpendicular to the plane (along the spatial $z$ axis) that hosts the HES (i.e. the spatial $x-y$ plane), and therefore, $S_z$ serves as a good quantum number to label the HES. The dynamics of the helical edge can effectively be described in terms of a Dirac Hamiltonian of the form
\begin{equation}
 \mathcal{H}_{\rm QSH} = \int dx~\Psi^\dagger \mathcal{H} \Psi~~;~~ \mathcal{H} = -i\hbar({\bf a}_{\rm SO}\cdot{\bf \sigma})\partial_x,
\end{equation}
where $x$ denotes the spatial coordinate along an edge, ${\bf a}_{\rm SO}$ is the SO field orienting along the spatial $z$-axis: ${\bf a}_{\rm SO}=v_F(0,0,1)$; $v_F$ being the Fermi velocity of the electrons on the edge and $\Psi\equiv(\psi_R~~\psi_L)^T$ denotes the annihilation operator for the right ($R$) and the left ($L$) moving electrons (they can equivalently be labeled by $\uparrow$ or $\downarrow$).

In general, the SO field along the edge can orient along any arbitrary direction destroying the conservation of $S_z$. It is only the time-reversal symmetry that suffices to preserve the HES implying that the spin rotation symmetry about the $z$ axis can be broken without influencing the topology of the bulk. Such freedom of tuning the SO field direction allows for the possibility of the generation of {\it spin Berry} (SB) phase~\cite{BERARD2006190,Murakawa1490} that can arise because of spin dynamics of the electron in addition to the dynamical phase produced due to its propagation along the edge. This phase can be understood as Aharonov-Bohm (AB) effect on the Bloch sphere~\cite{aharonov59, berry1987adiabatic}, and hence, is referred to as spin AB effect. Many authors, in the last few decades, have explored the presence of such phase appearing in the context of mesoscopic transport set-ups~\cite{meir1989universal, loss&goldbart, stern1992, Aronov1993, wadhawan2018interferometry}.  

There have been recent theoretical proposals which have explored the possibility of probing the helical nature of the edge state in transport set-up~\cite{corner}. In this paper, we are particularly interested in interferometric signatures and manifestation of helical nature of the edge state. In this context, Maciejko et al.~\cite{maciejko2010spin} studied the possibility of building a spin transistor in a AB ring built into a QSH state which is sandwiched between two ferromagnetic leads. They showed that it is possible to control spin of the electron on the edge via the AB flux resulting in spin AB effect. However, they assumed a uniform SO coupling along the edge maintaining the conservation of $S_z$. 

\begin{figure}
\centering
\includegraphics[width=1.0\columnwidth]{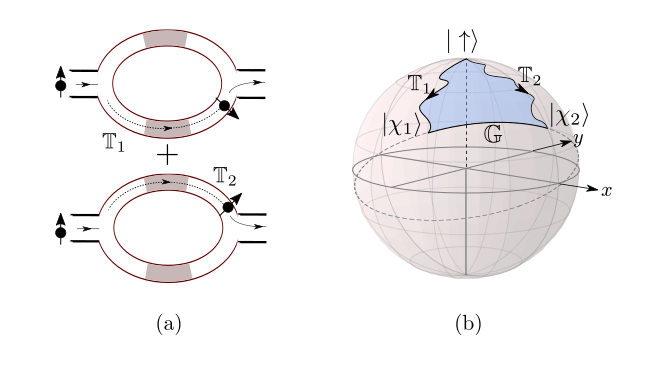}
\caption{
(a) Schematic of the two-path interferometer to realize the spin AB effect. The two interfering paths mentioned in the main text are depicted as ${\mathbb{T}}_{1}$ and ${\mathbb{T}}_{2}$ and the grey shades represent the SO field-active regions. (b) The trajectories corresponding to ${\mathbb{T}}_{1}$ and ${\mathbb{T}}_{2}$ in (a) are cast on the Bloch sphere. The geodesic $\mathbb{G}$ connects the end points forming a closed loop surrounding the blue shaded region.}
\label{fig:figg1}
\end{figure}

In contrast to their work, we study a complementary situation where the electron spin on the edge is itself undergoing a nontrivial variation along the edge due to the presence of a nonuniform SO field on the edge, hence, destroying the $S_z$ conservation. We discuss the minimal scenario where such a variation could lead to a fictitious flux induced by the spin Berry phase. 

Earlier theoretical study also predicted evidence of quantized geometric phase~\cite{SBP1} of $\pi$ where transport across a Fabry-Perot interferometer is studied using a double quantum point contact (QPC) geometry in a QSH state. Our study generalizes all such results to the case of nonquantized geometric phase. Effects due to gate induced doping of the edge state resulting from the application of an electrical field along a finite patch of the edge state have also been studied~\cite{SBP2}. This study exploited the gate controlled dynamical phase for tuning the interference signal in QSH interferometer and was insensitive to the SO interaction induced by the electric field of the applied gate. Therefore, it could not distinguish the $S_z$ nonconserving case from the conserving one. Addressing the former is the focus of our study. 

Noninterferometric signatures of scattering of electrons from a SO barrier induced by application of a local gate voltage have also been studied in the context of interacting helical edge state~\cite{Roni}. However, in that work also, the primary focus was on a uniform SO barrier and the possibility of realizing a nonzero geometric phase arising from the variation of the SO field along the edge was not considered. 

To gain insight into the {\it generation and detection} of spin Berry phase in an interferometer set-up, let us consider a standard two-path interferometer~\cite{aharony2002phase, ji2003electronic} as a prototype. Let us further assume that the interferometer arms are endowed with the possibility of rotating the electron spin due to the presence of SO coupling~\cite{datta1990electronic} in the arms of the interferometer as it traverses through the respective arms of the interferometer. In this paper, we will discuss specific models of SO-coupled Hamiltonians that serve as the {\it necessary and sufficient} requirement for inducing the rotation of the spin that allows it to acquire a finite SB phase in its closed loop journey around the interferometer. For further illumination, the following scenario would be useful to consider. Let us assume an electron with spin $\vert\uparrow\rangle$ entering the interferometer from the left lead [Fig.~\ref{fig:figg1} (a)] and its wavefunction simultaneously leaking into the upper and lower arms with respective quantum mechanical amplitudes. As the amplitudes propagating along the upper and the lower arms could generically suffer different history of the SO field, the incident spinor would evolve into $\vert\chi_{1}\rangle$ in the upper arm and $\vert\chi_{2}\rangle$ in the lower arm that trace out two independent trajectories (labeled ${\mathbb{T}}_1$ and ${\mathbb{T}}_2$ starting from the same point corresponding to the incident state $\vert\uparrow\rangle$ on the Bloch sphere [Fig.~\ref{fig:figg1} (b)]. Following Ref.~\onlinecite{berry1987adiabatic}, we arrive at the conclusion that the resulting interference pattern will depend on an extra phase factor which is given by half the solid angle subtended at the center by the closed area surrounded by ${\mathbb{T}}_1$, ${\mathbb{T}}_2$ and the geodesic~\cite{sam} ${\mathbb{G}}$ connecting $\vert\chi_{1}\rangle$ and $\vert\chi_{2}\rangle$  on this Bloch sphere [Fig.~\ref{fig:figg1} (b)]. This phase is the same as the AB phase accumulated by an electron while traversing once around the periphery of the above defined area ($\mathcal{A}\{{\mathbb{T}}_1$,${\mathbb{T}}_2$,${\mathbb{G}}\}$) on the surface of a unit sphere while a  monopole of strength half sits at the center of this sphere~\cite{berry1984}. The tunability of the orientation of the spin would result in modulations of the phase which manifest as oscillations in the current through the interferometer and can be visualized as stretching and shrinking the above mentioned area on the Bloch sphere by changing ${\mathbb{T}}_1$ or ${\mathbb{T}}_2$ or both in a controlled manner. This discussion provides us with a clear picture regarding the generation and detection of a finite SB phase in such a two-path interferometer geometry.

The rest of the paper is organized as follows: In section~\ref{secone}, we discuss the minimal scenario leading to a finite SB phase on the helical edge state resulting either from an intrinsic SO interaction of the spin Hall state or due to the application of an external electric field on the edge. In section~\ref{sectwo}, we calculate the transfer matrix for the situation which corresponds to minimal scenario for hosting a finite SB phase. Then in section~\ref{secthree}, we show that a two-terminal transport set-up involving spin-polarized leads provides a clear signature of the SB phase for different possible orientations of the spin polarization of the leads. {\color{black}{Finally in section~\ref{secfour}, we simulate a lattice version of the interferometer involving a modified BHZ model of QSH state using KWANT package~\cite{groth2014kwant} which demonstrates the essential physics of the resonances and antiresonances. We summarize the results and conclude in section~\ref{secfive}.}} 

\begin{figure*}
\centering
\includegraphics[width=2.0\columnwidth]{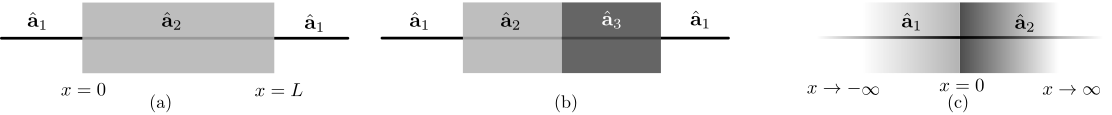}
\caption{(a) SO field configuration leading to a topological ($0$ or $\pi$) SB phase. (b) A minimal condition on the SO field configuration leading to the accumulation of a finite SB phase. (c) An interface between two distinct SO barriers considered to construct the transfer matrix as discussed in the main text.}
\label{fig:figg2}
\end{figure*}

\begin{figure}
\centering
\includegraphics[width=1.0\columnwidth]{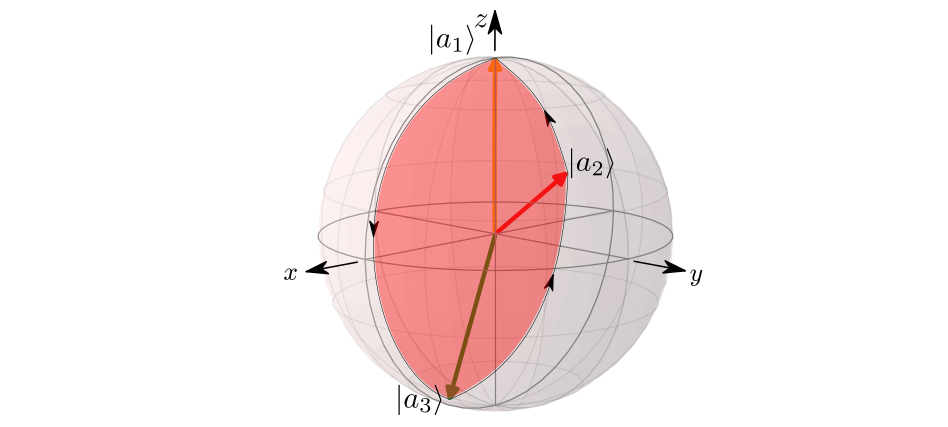}
\caption{
The SB phase is proportional to the area of the triangle formed on the Bloch sphere by the spinors $|a_1\rangle$, $|a_2\rangle$, and $|a_3\rangle$ corresponding to the SO field configuration shown in Fig.~\ref{fig:figg2} (b).}
\label{fig:figg3}
\end{figure}


\section{Scattering through spin-orbit barriers and spin Berry  phase}\label{secone}

In this section, we will discuss the possibility for an electron to accumulate a finite SB phase as it traverses through a nonuniform SO region which is embedded in otherwise uniform helical edge state. 
{\color{black}{This SO region can be spontaneously generated in the system without breaking the time-reversal symmetry either in a uniform fashion or fragmented into multiple small patches each hosting the SO field oriented in an arbitrary direction owing to the $S_z$-symmetry breaking in the bulk as will be demonstrated in section~\ref{secfour}. However, to perform analytic calculations, we, from now on, would consider specific profiles of the SO fields along the edges.}}

{\color{black}{We take the following Hamiltonian for the edge state (extended from $x=-\infty$ to $x=+\infty$, $x$ representing an intrinsic one-dimensional coordinate along the edge)}}
\begin{equation}
 H_{\rm SO} = - \frac{i}{2}\hbar \{{\bf a}(x),\partial_x\}\cdot\sigma,
 \label{Hso}
\end{equation}
where the spatial profile of the SO field ${\bf a}(x)$ is
\begin{equation}
 {\bf a}(x) = [1-\Theta(x)+\Theta(x-L)]{\bf a}_1 + [\Theta(x) -\Theta(x-L)]{\bf a}_2.
\end{equation}
Here $\Theta(x)$ denotes the Heaviside step function. To be specific, we consider a situation where the vector ${\bf a}_1= \vert {\bf a}_1\vert{\hat z}$ corresponds to a uniform SO field which is pointing along $z$-axis while ${\bf a}_2$, which is extended from $x=0$ to $x=L$ [Fig.~\ref{fig:figg2} (a)], can point in a direction different from ${\bf a}_1$ and can also have spatial variation. This finite patch of ${\bf a}_2$ can be thought of as a barrier. 

We consider a simplest possible situation where ${\bf a}_2$ represents a vector which is constant in space but is pointing in a direction different from ${\bf a}_1$ and further assume WLOG ${\bf a}_2= \vert {\bf a}_2\vert{\hat x}$. Note that, though the SO field is constant along the barrier, the electron spin undergoes a drastic change as it enters and exits the barrier when incident either from the left or from the right side of the barrier. Hence, {\it apriori} it is not clear if such situation would lead to a finite SB phase or not.

When an electron is incident on the SO barrier from the left ($x<0$) [Fig.~\ref{fig:figg2} (a)], its spin will initially point along the $z$-axis (north pole on the Bloch sphere) but once it enters the barrier it will reorient itself along the $x$-axis and again when it exits, the spin will rotate back to the $z$-axis (north pole). This implies that the trajectory of electron spin on the Bloch sphere traces a single curve (geodesic path) running from the north pole to the equator when it enters the barrier and then runs back exactly along the same path during its return journey when it exits. Hence, the trajectory of the spin state on the Bloch sphere encloses zero area during its close loop journey starting from and ending at the north pole and so, a zero SB phase accumulation is expected for a constant SO barrier. 

For accumulation of a finite SB phase we surely need a SO barrier which has a variation of the orientation of the SO field along the length of the barrier. The cases of nonzero SB phase can be categorized as follows: \\

(a) quantized SB phase of $\pi$,  \\

(b) nonquantized SB phase varying between $0$ and $2\pi$. \\

\noindent
A quantized value of SB phase can be generated by means of engineering the following SO barrier. The SO field $\hat{{\bf a}}(x)$ on the edge is chosen to be such that it, inside the barrier [Fig.~\ref{fig:figg2} (a)], rotates along the edge where the rotation is parameterized by a space dependent  monotonically increasing angle $\theta_x$ such that $\hat{{\bf a}}(x)= (\sin \theta_x,0,\cos \theta_x)$ while outside the barrier it is $\hat{{\bf a}}(x)= (0,0,1)$ implying $\theta_{x=0-}=\theta_{x=L+}=0$. Then it can be shown that 
\begin{equation}
  \phi_{\rm SB} = \left \{
  \begin{aligned}
    &0, && \text{if}\  \theta_{x=L-} < \pi\\
    &\pi && \text{if}\ \theta_{x=L-} > \pi,
  \end{aligned} \right.
  \label{topoSB}
\end{equation} 
where $\theta_{x=L-}$ specifies the orientation of the SO field right before exiting the barrier at $x=L$.  
In the case when $\phi_{\rm SB}=0$, the trajectory of the incident electron spin on the Bloch sphere traces a closed loop path along the great circle defined by the intersection of the $x$-$z$ plane and the Bloch sphere which goes back and forth on the Bloch sphere without encircling the center. This trajectory on the Bloch sphere is similar to the one for the case of constant SO barrier discussed previously. For the case of $\phi_{\rm SB}=\pi$, the electron spin on the Bloch sphere winds the great circle once as the electron traverses through the barrier and exits. This demonstrates the topological nature of this phase.  
{\color{black}{Here we would like to point to an important aspect of our work in distinction to the one reported in Ref.~[\onlinecite{SBP1}]. 
In Ref.~[\onlinecite{SBP1}], the authors considered the very special case of $S_z$ nonconservation via Rashba-type SO interactions switched on only at the bottlenecks of the interferometer geometry such that the variation of the spin is restricted to lie on the $y-z$ plane leading to a quantized SB phase. We have shown that even if the variation of spin is planar e.g. as considered by them, the corresponding SB phase {\it may or may not} be $\pi$ depending on the details of the planar variation. This possibility is summarized in Eq.~\ref{topoSB}. In short, we note that inverting the SO field at the two bottlenecks of the interferometer proposed by Ref.~[\onlinecite{SBP1}] is not the only way to generate a topological SB phase.}}   

Now we will discuss the minimal variation of the SO field within the barrier required to give rise to a finite nonquantized SB phase. We need to find a configuration of the SO field which will lead to closed loop trajectory of the electron spinor on the Bloch sphere enclosing a finite area as the electron enters and exits the SO barrier. This can be achieved if the barrier can be subdivided into two regions with their respective SO vectors pointing along $\hat{{\bf a}}_2$ first and then $\hat{{\bf a}}_3$ (starting from the left) which should be distinct from each other and also mutually distinct from $\hat{{\bf a}}_1$ [Fig.~\ref{fig:figg2} (b)]. The journey of the electron across such barrier, when incident from the left, can be mapped to the journey of the electron spinor on the Bloch sphere which is as follows. The incident spinor which is pointing to the north pole ($\hat{{\bf a}}_1$ being along $z$-axis) first moves to a point (call it $N_1$) on the surface of the Bloch sphere corresponding to the direction of $\hat{{\bf a}}_2$ along a geodesic path connecting the north pole and $N_1$. Then, as the electron further moves from region 1 to region 2 inside the barrier, its spinor moves from point  $N_1$ to point $N_2$ along the geodesic path connecting $N_1$ and $N_2$ on the Bloch sphere, where $N_2$ is the point on the surface of the Bloch sphere corresponding to the direction of $\hat{{\bf a}}_3$. Finally, when the electron leaves the barrier, the electron spinor moves back to the point corresponding to $\hat{{\bf a}}_1$ along a geodesic starting from $N_2$, hence, forming a spherical triangle on the  Bloch sphere. The SB phase accumulated by the electron in this journey will be given by half the solid angle ($\phi_{\rm SB}=\mathcal{A}/2$) subtended by the area $\mathcal{A}$ of the spherical triangle whose vertices are formed by the spinors $|a_1\rangle$, $|a_2\rangle$, and $|a_3\rangle$ (Fig.~\ref{fig:figg3}) which are the ``up" eigenstates with eigenvalue $+1$ of the corresponding $\hat{\bf a}\cdot\sigma$ Hamiltonian. The expression of $\mathcal{A}$ is given by~\cite{eriksson1990measure}
\begin{equation}
 \mathcal{A} = 2\tan^{-1} \frac{|\hat{{\bf a}}_1\cdot(\hat{{\bf a}}_2 \times \hat{{\bf a}}_3)|}{1+\hat{{\bf a}}_1\cdot\hat{{\bf a}}_2+\hat{{\bf a}}_2\cdot\hat{{\bf a}}_3+\hat{{\bf a}}_3\cdot\hat{{\bf a}}_1}~.
 \label{Bsphere}
\end{equation}   
In what follows, we will provide a derivation of this result using the transfer matrix method for reasons to be clear afterwards.


\section{SB phase and  transfer matrix}\label{sectwo}

In scattering problems, the computation of $\phi_{\rm SB}$ can be formulated in terms of transfer matrices that directly connect to the transport properties of the system concerned. For a generic profile of the SO field ${\bf a}(x)$, the Schr\"{o}dinger equation $H_{\rm SO}\Psi=E\Psi$ has solutions of the form $\Psi(x_2) = T_{x_2,x_1} \Psi(x_1)$ where the transfer matrix~\cite{timm2012transport} is given by
\begin{equation}
 T_{x_2,x_1} = \mathcal{P}_x {\rm Exp}\bigg[\int_{x_1}^{x_2} {\rm d}x~\frac{{\bf a}\cdot\sigma}{\hbar|{\bf a}|^2}\big(E+\frac{i\hbar}{2}\partial_x{\bf a}\cdot\sigma\big)\bigg],
 \label{Tmat}
\end{equation}
where  $\mathcal{P}_x$ represents path-ordering (to derive Eq.~\ref{Tmat}, one needs to recast the Schr\"{o}dinger equation $H_{\rm SO}\Psi=E\Psi$ as $\partial_x \Psi=H_0(x)\Psi$ and use $T_{x_2,x_1}=\mathcal{P}_x {\rm Exp}\big[\int_{x_1}^{x_2} {\rm d}x~H_0(x)\big]$).\\

Now, let us consider a situation corresponding to an abrupt change of the SO field at $x=0$ [Fig.~\ref{fig:figg2} (c)] which can be modeled as,  
\begin{equation}
 {\bf a}(x) = [1-\Theta(x)]{\bf a}_1+\Theta(x){\bf a}_2,
\end{equation}
where ${\bf a}_1$ and ${\bf a}_2$ are two constant SO fields with distinct directions in region 1 ($x<0$) and 2 ($x>0$) respectively. 
Substituting this expression of ${\bf a}(x)$ into Eq.~\ref{Tmat}, we obtain a matching condition between the spinors on the two sides of the interface which reads $\psi(0^+)=T_{21}\psi(0^-)$ where $T_{21}$ denotes a transfer matrix from the region of ${\bf a}_1$ to the region of ${\bf a}_2$ [in Fig.~\ref{fig:figg2} (c)] and is of the form
\begin{equation}
 T_{21} = \sqrt{\frac{|{\bf a}_1|}{|{\bf a}_2|}} {\rm Exp}\big[i\,\theta_{21}\,{\hat{\bf D}}_{21}\cdot{\sigma}\big],
 \label{Tmat21}
\end{equation}
where
\begin{align}
 {\bf D}_{21} &= {\bf a}_2\times{\bf a}_1,~~\text{and}~~\tan(2\theta_{21})=|{\bf D}_{21}|/({\bf a}_2\cdot{\bf a}_1).
\end{align}
Note the operator $T_{21}$ is, in general, a nonunitary operator unless $\vert{\bf a}_1\vert=\vert{\bf a}_2\vert$. A minimal set-up required for obtaining a nonzero SB phase corresponds to an array of such interfaces between distinct SO fields and needs to be constructed such that the electron spin, in successive steps, encounters the SO fields as  $\hat{{\bf a}}_1\rightarrow \hat{{\bf a}}_2\rightarrow \hat{{\bf a}}_3\rightarrow \hat{{\bf a}}_1$ as noted previously and shown in Fig.~\ref{fig:figg2} (b). The net transfer matrix in this process is remarkably an unitary operator of the form
\begin{align}
 \mathcal{T} &= \prod_{ij} T_{ij} = T_{13}T_{32}T_{21}  \nonumber \\
 & = {\rm Exp}\big[i\theta_{13}{\hat{\bf D}}_{13}\cdot{\sigma}\big]{\rm Exp}\big[i\theta_{32}{\hat{\bf D}}_{32}\cdot{\sigma}\big]{\rm Exp}\big[i\theta_{21}{\hat{\bf D}}_{21}\cdot{\sigma}\big],
 \label{Tprod}
\end{align}
irrespective of the magnitudes of ${\bf a}_1$, ${\bf a}_2$, and ${\bf a}_3$. 
The SB phase acquired by the electron as it goes once around the circle defined by $\hat{{\bf a}}_1\rightarrow \hat{{\bf a}}_2\rightarrow \hat{{\bf a}}_3\rightarrow \hat{{\bf a}}_1$ [Fig.~\ref{fig:figg2} (b)] is given by $\phi_{\rm SB}={\rm arg}[\langle n_{\uparrow1}|T|n_{\uparrow1}\rangle]$ where the subscript 1 denotes that the spinor whose evolution is concerned is an eigenstate of $H_{\rm SO}$ in Eq.~\ref{Hso} with ${\bf a}={\bf a}_1$ and $\uparrow$ represents the spin state of the electron aligned with the local SO field while it is moving along $\hat{{\bf a}}_1\rightarrow \hat{{\bf a}}_2\rightarrow \hat{{\bf a}}_3\rightarrow \hat{{\bf a}}_1$ ($\downarrow$ would correspondingly represent the spin of the electron antialigned with the local SO field while moving in the opposite direction).

To obtain an explicit expression of the SB phase, Eq.~\ref{Tprod} can be re-expressed in a compact form as 
\begin{align}
 \mathcal{T} \equiv {\rm Exp}\big[i\,\alpha\,\hat{\mathcal{K}}\,\cdot\,{\sigma}\big],
 \label{TTprod}
\end{align}
where 
\begin{align}
 \cos\alpha &= \cos\theta_{13}\cos\theta_{32}\cos\theta_{21} \nonumber \\
 & - ({\hat{\bf D}}_{13}\cdot{\hat{\bf D}}_{32})\sin\theta_{13}\sin\theta_{32}\cos\theta_{21} \nonumber \\
 & - ({\hat{\bf D}}_{32}\cdot{\hat{\bf D}}_{21})\cos\theta_{13}\sin\theta_{32}\sin\theta_{21} \nonumber \\
 &-({\hat{\bf D}}_{21}\cdot{\hat{\bf D}}_{13})\sin\theta_{13}\cos\theta_{32}\sin\theta_{21}  \nonumber \\
 & + [{\hat{\bf D}}_{13},{\hat{\bf D}}_{32},{\hat{\bf D}}_{21}]\sin\theta_{13}\sin\theta_{32}\sin\theta_{21},
\label{sangle}
\end{align}
and 
\begin{align}
 \hat{\mathcal{K}}\sin\alpha 
 &= {\hat{\bf D}}_{13}\sin\theta_{13}(\cos\theta_{32}\cos\theta_{21} \nonumber \\
 &-{\hat{\bf D}}_{32}\cdot{\hat{\bf D}}_{21}\sin\theta_{32}\sin\theta_{21}) \nonumber \\
 &+ {\hat{\bf D}}_{32}\sin\theta_{32}(\cos\theta_{13}\cos\theta_{21} \nonumber \\
 &+{\hat{\bf D}}_{13}\cdot{\hat{\bf D}}_{21}\sin\theta_{13}\sin\theta_{21}) \nonumber \\
 &+{\hat{\bf D}}_{21}\sin\theta_{21}(\cos\theta_{32}\cos\theta_{13} \nonumber \\
 &-{\hat{\bf D}}_{32}\cdot{\hat{\bf D}}_{13}\sin\theta_{32}\sin\theta_{13}) \nonumber \\
 &+({\hat{\bf D}}_{21}\times{\hat{\bf D}}_{13})\sin\theta_{13}\cos\theta_{32}\sin\theta_{21} \nonumber \\
 & - ({\hat{\bf D}}_{13}\times{\hat{\bf D}}_{32})\sin\theta_{13}\sin\theta_{32}\cos\theta_{21} \nonumber \\ 
 &- ({\hat{\bf D}}_{32}\times{\hat{\bf D}}_{21})\cos\theta_{13}\sin\theta_{32}\sin\theta_{21}. 
\label{uvec}
\end{align}
A straightforward but lengthy algebra leads to the following expression of $\alpha$ given by 
\begin{align}
 \tan\alpha = \frac{|\hat{{\bf a}}_1\cdot(\hat{{\bf a}}_2 \times \hat{{\bf a}}_3)|}{1+\hat{{\bf a}}_1\cdot\hat{{\bf a}}_2+\hat{{\bf a}}_2\cdot\hat{{\bf a}}_3+\hat{{\bf a}}_3\cdot\hat{{\bf a}}_1}~.
 \label{tri}
\end{align}
Here, $\alpha$ has a natural interpretation as the SB phase owing to the fact that ${\mathcal{\hat K}}$ is collinear with ${\hat{\bf a}}_1$ and $\alpha$ appears as an overall phase in Eq.~\ref{TTprod}. The unit vector ${\mathcal{\hat K}}$ will be parallel to ${\hat{\bf a}}_1$ (${\mathcal{\hat K}} = {\hat{\bf a}}_1$) when the sense of circulation of the electron spinor represented on the Bloch sphere is clockwise. On the other hand, if the sense of the circulation is anticlockwise, then ${\mathcal{\hat K}}$ will be antiparallel to ${\hat{\bf a}}_1$ (${\mathcal{\hat K}} = -{\hat{\bf a}}_1$).The explicit derivation of Eq.~\ref{tri} is given in Appendix~\ref{appA}. Note the expression in Eq.~\ref{Bsphere} is exactly the same as Eq.~\ref{tri} with ${\bf \hat z}\rightarrow {\hat{\bf a}_1}$, ${\bf \hat a_1}\rightarrow {\bf \hat a_2}$ and ${\bf \hat a_2}\rightarrow {\bf \hat a_3}$ and $\alpha$ being identified as half the solid angle subtended by the area of a spherical triangle (shown in Fig.~\ref{fig:figg3}) on the Bloch sphere. 

{\color{black}{It is to be noted that the derivation of Eq.~\ref{tri} is crucial for it provides us a firm ground to establish a connection between the intuitive picture of drawing trajectories on the Bloch sphere and the electron transport quantified via the path-ordered product of transfer matrices, or in other words, the connection between the evolution of the electron in physical space and the trajectory of the corresponding spin on the Bloch sphere.
}}


\section{SB phase and its interferometric manifestation}\label{secthree}

Here we study a minimal two-terminal transport set-up (for a sketch see Fig.~\ref{fig:figg5}) which could lead to the detection of a finite SB phase. Our proposed set-up is a ring (of circumference L) representing an isolated closed edge (see Fig.~\ref{fig:figg5}) which is tunnel-coupled to two polarized leads $L_1$ and $L_2$ at the point P1 and P2 as shown in Fig.~\ref{fig:figg5}. As discussed earlier in the previous section, a finite SB phase requires the presence of a spatially varying SO field and a minimal scenario demands for the presence of at least three distinct directions of the SO field along the edge [as in Fig.~\ref{fig:figg2} (b)]. Hence we consider a model where the SO field configuration along the edge is taken to be such that the entire ring is covered by three successive patches of SO field pointing along ${\bf a}_1$, ${\bf a}_2$ and ${\bf a}_3$. We have chosen ${\bf \hat{a}}_1=(0,0,1)$ for calculational convenience. Note that though our calculations are done for a model with sudden jump between different SO field directions, our qualitative results remain even if we replace our situation with another situation where these three regions are connected such that the vectors ${\bf a}_1$, ${\bf a}_2$ and ${\bf a}_3$ go smoothly on to one another. It is important to take note of this point as the primary aim of this paper is to address an edge state where $S_z$ is not conserved, i.e., the orientation of the spin along the edge is smoothly varying over space.  

\subsection{Interferometry with polarized leads}

\begin{figure}
\centering
\includegraphics[width=1.0\columnwidth]{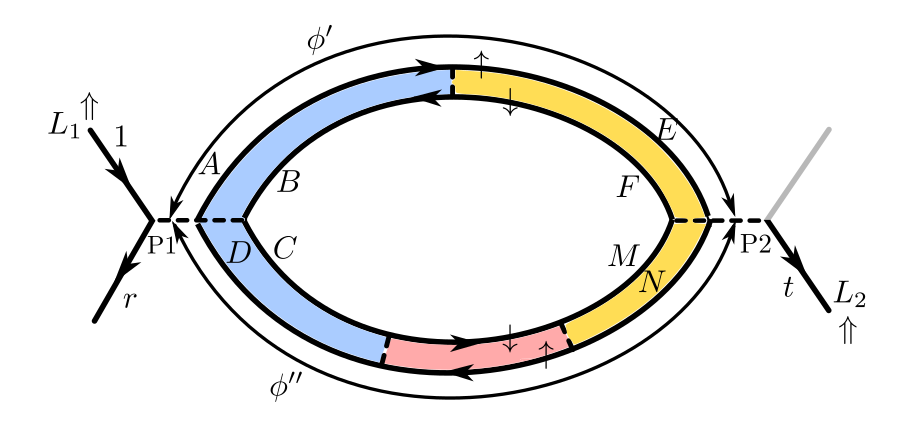}
\caption{Sketch of the ${\bf a}_1-{\bf a}_2-{\bf a}_3$ model with two leads connected at P1 and P2 is shown. The SO fields in the blue, yellow, and red region are respectively given by ${\bf a}_1$, ${\bf a}_2$, and ${\bf a}_3$. The respective amplitudes of propagation are denoted at the tunnel-junctions P1 (injecting) and P2 (receiving). The phases $\phi'$ and $\phi''$ are the dynamical phases from P1 to P2 along the upper and lower arm respectively and the total length of the arms is $L$. The spinors of the leads ($L_{1,2}$) are denoted by $\Uparrow$ while that of the HES are denote by $\uparrow,\downarrow$.}
\label{fig:figg5}
\end{figure}

The schematic of the proposed set-up is given in Fig.~\ref{fig:figg5} where the closed edge state is subdivided into three parts such that the SO fields in the blue, yellow, and red region are specified by three distinct vectors ${\bf a}_1$, ${\bf a}_2$, and ${\bf a}_3$ respectively. The Hamiltonian for the closed edge is provided in Eq.~\ref{Hso} with a given spatial profile of ${\bf a}(x)$ subjected to periodic boundary condition. We have considered two tunnel-coupled spin-polarized leads ($L_1$ and $L_2$) attached to the closed edge where $L_1$ is coupled to a point P1 in the blue region and $L_2$ is coupled to a point P2 in the yellow region (the lead positions are arbitrary and can be in any one/two of the three regions; we, for instance, consider the case when the two leads are placed in two different regions). We model the leads by spin-polarized chiral edge states with linear dispersion. This way, owing to the linear dispersion, the transport is influenced only by the direction of spin polarization of the lead electrons and not by the density of states of the leads. The form of the lead Hamiltonian (for lead $L_I$, $I=1,2$) can be taken to be a chiral mode (right moving with Fermi velocity $v_F$) and is given by 
\begin{equation}
 \mathcal{H}_{L_I} = -\imath \hbar v_F \int~{\rm d}x~ \psi_{\Uparrow I}^\dagger \partial_x \psi_{\Uparrow I},
 \label{Htip}
\end{equation}
where $\psi_{\Uparrow I}^\dagger$ creates an electron in lead $L_I$ with spinor $|n_{\Uparrow I}\rangle$. The Hamiltonian which defines a tunnel-junction at $x=x_{P_I}$ corresponding to the tunnel-coupling between lead $L_I$ and the corresponding edge takes a form given by 
\begin{equation}
 \mathcal{H}_{\rm T}^{(I)} = {\Gamma_I}\int~{\rm d}x~\delta(x-x_{P_I}) \sum_{\alpha=\uparrow,\downarrow} \big\{ \Upsilon_{\alpha j,\Uparrow I}\psi_{\alpha j}^\dagger \psi_{\Uparrow I} + {\rm h.c.} \big\},
 \label{htun}
\end{equation}
where $\psi_{\alpha j}^\dagger$ represents the creation operator for electrons in the HES with spinor $|n_{\alpha j} \rangle$ specified by the Hamiltonian in Eq.~\ref{Hso} with ${\bf a}={\bf a}_j$ and $\Upsilon_{\alpha j,\Uparrow I}=\langle n_{\alpha j}|n_{\Uparrow I}\rangle$; ${\Gamma_I}$ represents the tunneling strength at the tunnel-junction between lead $L_I$ and the (local) edge.

Now we set up the calculation of the transmission amplitude through the ring (Fig.~\ref{fig:figg5}). We consider a scattering problem where an electron is incident from lead $L_1$ and transmitted into lead $L_2$. This problem can be split into three different scattering problems which are finally connected to one another via boundary conditions as follows:\\
a) \textit{Scattering at P1} :\\
The scattering at point P1 can be reduced to a scattering between three incoming and three outgoing chiral edges at point P1. The incoming and the outgoing amplitudes at P1 are connected via scattering matrix~\cite{lesovik2011scattering} $S_1$ as
\begin{equation}
 \begin{pmatrix}
  A & C & r 
 \end{pmatrix}^T
= S_1
 \begin{pmatrix}
  D & B & i_1 
 \end{pmatrix}^T,
 \label{scatmat1}
\end{equation} 
where $i_1$($=1$) and $r$ are the plane wave amplitudes of the incident and the reflected wave in lead $L_1$ at P1. The incoming and outgoing amplitudes in the helical edge at P1 are given by $A$, $B$, $C$, and $D$.\\
b) \textit{Scattering at P2} :\\
Similarly, at point P2, the incoming and outgoing amplitudes are connected via scattering matrix $S_2$ as
\begin{equation}
 \begin{pmatrix}
  N & F & t 
 \end{pmatrix}^T
= S_2
 \begin{pmatrix}
  E & M & i_2
 \end{pmatrix}^T.
 \label{scatmat2}
\end{equation}
where $E$, $F$, $M$, and $N$ are the incoming and outgoing amplitudes in the ring and $t$ (transmission amplitude) is the outgoing amplitude in lead $L_2$. The incoming amplitude $i_2$ is zero as no incidence is considered in lead $L_2$. \\
c) \textit{Connecting the amplitudes inside the ring via transfer matrices given in  Eq.~\ref{Tprod}}: \\
Now to implement the matching conditions for the various amplitudes inside the closed ring, let us divide the ring into two parts as in Fig.~\ref{fig:figg5}: (i) the upper arm, where the journey of the electron ($\uparrow$) starting from point P1 $\rightarrow$ P2 in clockwise sense accumulates a dynamical phase of $\phi^{\prime}$ while the geometric phase (if any) is naturally embedded inside the transfer matrix, (ii) the lower arm, where the journey of the electron ($\uparrow$) starting from point P2 $\rightarrow$ P1 in the clockwise sense accumulates a dynamical phase of $\phi^{\prime\prime}$ while again the geometric phase (if any) is naturally embedded inside the transfer matrix. These phases are incorporated  into the problem via the following boundary conditions for the upper arm: 
\begin{align}
E &= \langle n_{\uparrow 2}| T_{21} |n_{\uparrow 1} \rangle e^{i \phi^{\prime}} A
\nonumber \\
B &= \langle n_{\downarrow 1}| T_{12} |n_{\downarrow 2} \rangle e^{i \phi^{\prime}} F,
\label{sticha}
\end{align}
while for the lower arm, they are given by,
\begin{align}
M &= \langle n_{\downarrow 2}| T_{23} T_{31} |n_{\downarrow 1} \rangle e^{i \phi^{\prime\prime}} C
\nonumber \\
D &= \langle n_{\uparrow 1}| T_{13} T_{32} |n_{\uparrow 2} \rangle e^{i \phi^{\prime\prime}} N,
\label{stichb}
\end{align}
where $|n_{\alpha i}\rangle$ ($\alpha=\uparrow/\downarrow$) represents the eigenstate with $\pm1$ eigenvalue of ${\bf \hat{a}}_i\cdot\sigma$ ($\uparrow \leftrightarrow +1$, $\downarrow \leftrightarrow -1$). 
Finally, these three steps (a), (b), and (c) together provide the transmission amplitudes ($t$) of the system whose explicit forms are given below. Now, three distinct physical scenarios can be realized depending upon the relative orientations of the spin polarization of the leads with respect to the orientations of the local SO fields of the edge to which the leads are being tunnel-coupled:\\

(1) Both leads local parallel $:-$ The spin polarization axes of both lead $L_1$ and $L_2$ are parallel to the vectors ${\bf a}_1$ and ${\bf a}_2$ respectively,  \\

(2) One of the leads local parallel $:-$ The spin polarization axis of lead $L_1$ is no more parallel to the vector ${\bf a}_1$, while that of lead $L_2$ is still taken to be parallel to the vector ${\bf a}_2$, \\

(3) Complete deviation from local parallel condition $:-$ Both the spin polarization axes of leads $L_1$ and $L_2$ are no more parallel to the vectors ${\bf a}_1$ and ${\bf a}_2$ respectively. \\

From now on, we will assume the Fermi velocity in the leads ($v_F$) and that in the ring to be the same implying $|{\bf a}_1|=|{\bf a}_2|=|{\bf a}_3|=v_F$. It is to be noted that such assumption does not influence the geometric phase aspect of the problem whatsoever as long as ${\bf \hat{a}}_1$, ${\bf \hat{a}}_2$, and ${\bf \hat{a}}_3$ are distinct. An explicit calculation of scattering matrices for mutual tunneling between different chiral edges is presented in Ref.~[\onlinecite{wadhawan2018interferometry}] by exploiting the equation of motion technique following which, here, we have calculated $S_{1,2}$ at tunnel-junctions P1 and P2 (Fig.~\ref{fig:figg5}) and also the transmission amplitudes for the three cases depicted above.

\subsubsection{First scenario: Both leads local parallel}

\begin{figure}
\centering
\includegraphics[width=1.0\columnwidth]{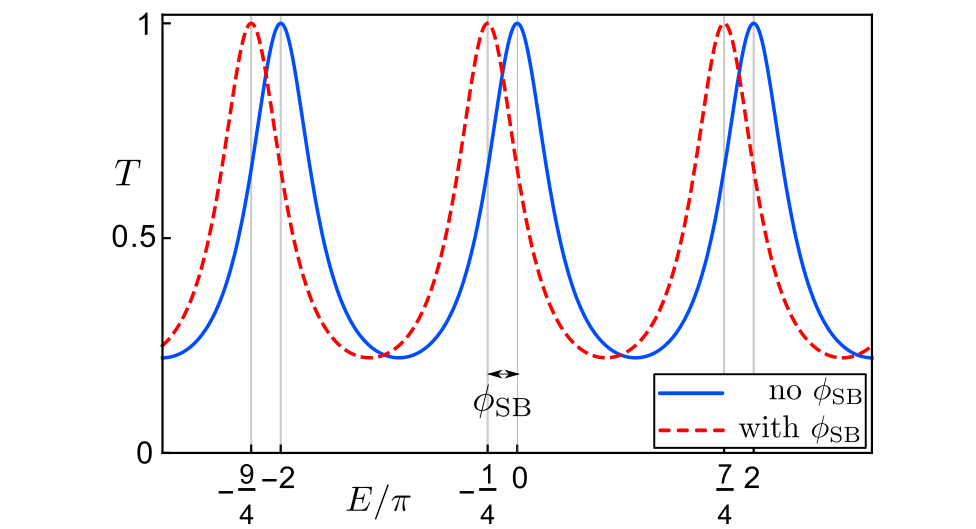}
\caption{{\bf First scenario:} Coherent oscillations observed in the transmission probability $T$ as a function of the incident energy $E$ ($=\hbar v_F\phi_D/L$) featuring resonance peaks at $E=2n\pi$ (as we take $\hbar=v_F=L=1$ for our calculations) which get shifted in presence of SB phase [$\phi_{\rm SB}=\pi/4$ in the plot with $\hat{\bf a}_1=(0,0,1)$, $\hat{\bf a}_2=(1,0,0)$, and $\hat{\bf a}_3=(0,1,0)$]. The tunneling strengths are taken as $\tilde{\Gamma}_1=\tilde{\Gamma}_2=1$.}
\label{fig:figg6}
\end{figure}

This case corresponds to the simplest possible situation the (spin-polarized) leads $L_1$ injects only clockwise moving ($\uparrow$) electrons into the ring as the spin polarization axis of the lead is taken to be parallel to the direction of the SO field at P1. Hence, the injected current flows only in the clockwise direction. In this case, the transmission amplitude from lead $L_1$ to lead $L_2$ can be straightforwardly obtained as 
\begin{align}
t = \frac{16 e^{i \phi^{\prime}} \tilde{\Gamma}_1 \tilde{\Gamma}_2}{-a + b e^{i \phi_D}},
\label{noSBphase}
\end{align}
where $a = (4+\tilde{\Gamma}_1^2) (4+\tilde{\Gamma}_2^2)$ , $b = (4-\tilde{\Gamma}_1^2) (4-\tilde{\Gamma}_2^2)$; $\tilde{\Gamma}_{1,2}=\Gamma_{1,2}/(\hbar v_F)$ are dimensionless parameters, $\Gamma_1$($\Gamma_2$) being the tunneling strength at the tunnel-junction P1 (P2) and $\phi_D$ ($=\phi'+\phi''$) is the total dynamical phase acquired by an electron in a full cycle of its journey along the edge (i.e. P1$\rightarrow$P2$\rightarrow$P1  traversing a length of $L$).

In presence of a finite SB phase, the transmission probability changes to $T(\phi_D)\rightarrow T(\phi_D + \phi_{SB})$ ($T={\vert t \vert}^2$) where the contribution due to the SB phase enters via matching conditions which depend on the transfer matrix as given in Eq.~\ref{sticha} and Eq.~\ref{stichb}. The interference pattern observed as a function of the incident energy $E=\hbar v_F \phi_D/L$ features resonance peaks at values of $E=\phi_D=2n\pi$ for integer $n$ (we have taken $\hbar=v_F=L=1$) when $\phi_{\rm SB}=0$ and the peaks are shifted such that at the resonance peaks, $\phi_D+\phi_{\rm SB}=2n\pi$ when $\phi_{\rm SB}\neq 0$. In particular, the case of quantized SB phase of $\pi$ results in a complete swapping of the maxima and minima of the transmission probability $T(E)$. Such special case of quantized SB phase will be very similar to the situation discussed in Ref.~[\onlinecite{SBP1}]. Hence, the shift of the maxima in transmission probability in the interference pattern would indicate the presence of a finite SB phase in our set-up.

To summarize, for the simplest possible scenario of local parallel leads, the interference pattern obtained as a function of incident energy of the electron is shown to be independent of the positions of P1 and P2. The SB phase, in this case, can be read off by measuring the shift of the resonance peaks in $T(E)$ (see Fig.~\ref{fig:figg6}). In particular, the resonance, which is expected at the Dirac point ($E = 0$), would shift due to the presence of a finite SB phase. Hence, as long as the identification of the Dirac point could be made by some independent experimental technique, the manifestation and quantification of the SB phase can be directly related to the shift of the resonance peak from the Dirac point.

\subsubsection{Second scenario: partial (one lead) deviation from local parallel condition: antiresonance}
\begin{figure}
\centering
\includegraphics[width=1.0\columnwidth]{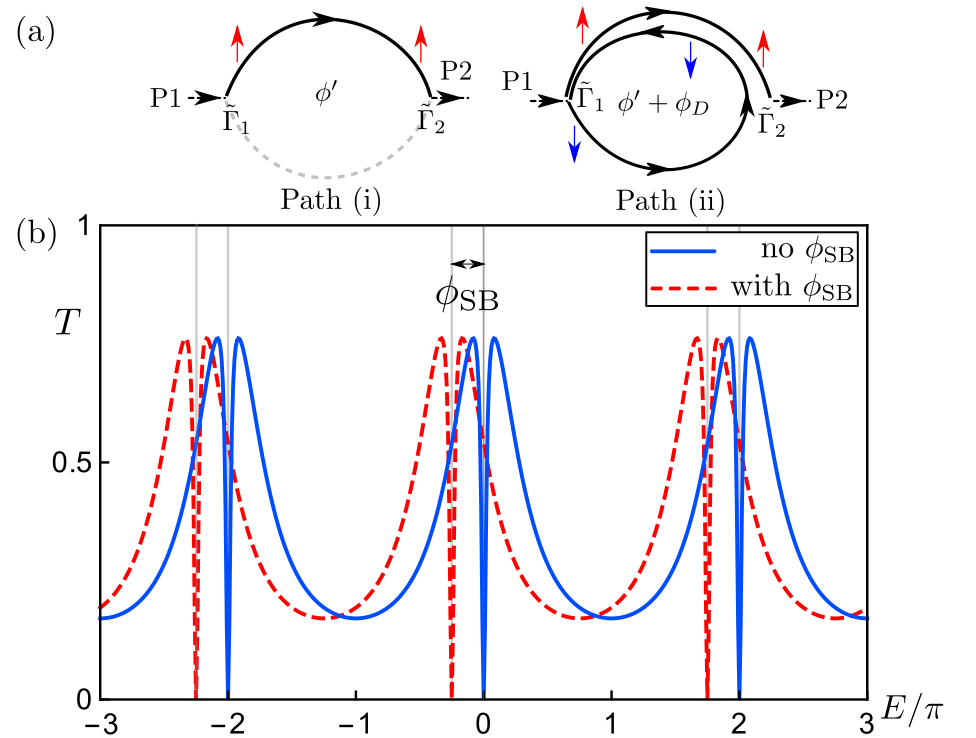}
\caption{{\bf Second scenario:} (a) The two distinct types of paths (described in the text) that lead to a destructive interference yielding antiresonances at $E=2n\pi$ when one of the polarized leads is deviated away from its local parallel configuration. (b) The resultant interference pattern in absence (solid) and presence (dashed) of the SB phase $\phi_{\rm SB}$ [$\phi_{\rm SB}=\pi/4$ in the plot with $\hat{\bf a}_1=(0,0,1)$, $\hat{\bf a}_2=(1,0,0)$, and $\hat{\bf a}_3=(0,1,0)$]. Lead $L_2$ is kept local parallel while lead $L_1$ is tilted by an angle of $\pi/3$ from $\hat{\bf a}_1$ keeping the azimuthal angle same. The tunneling strengths are taken as $\tilde{\Gamma}_1=\tilde{\Gamma}_2=1$.}
\label{fig:figg7}
\end{figure}

In this case, lead $L_1$ is no more ``local parallel" to the direction of the SO coupling at the tunnel-junction P1 and hence, it injects into both the clockwise and anticlockwise moving edge channels while the other lead $L_2$ can only absorb one particular chirality (right movers). As the electron traverses from $L_1$ to $L_2$, the two leading contributions to the transport can be attributed to the two distinct types of path and their interference [shown in Fig.~\ref{fig:figg7} (a)]. \\
(i) The first type of path [shown in Fig.~\ref{fig:figg7} (a) left] is related to the injection of a clockwise moving $\uparrow$-electron ($\uparrow$ with respect to the local SO field) at $L_1$ which has a finite probability amplitude to exit at $L_2$ after a direct traversal along the upper arm without going around the ring. Rest of the subsequent paths corresponds to the electron undergoing multiple rounds of circulation along the ring  before exiting. The important point to note here is the fact that, to exit, the electron must be in an ``up" state (with respect to the local SO field region that holds $L_2$) as $L_2$ is a local parallel lead receiving only the ``up" states. To ensure this, the electron circulating along the ring has two possibilities:- either it goes around the ring integer number of times as a clockwise mover ($\uparrow$-electron) without suffering spin-flip backscattering at $L_1$ via a second-order tunneling process (between the lead and the edge), or, if the electron suffers spin-flip backscattering at $L_1$, it must undergo such scattering even number of times so that it could return back to the up state before it could exit via lead $L_2$.
(ii) The second type of path [shown in Fig.~\ref{fig:figg7} (a) right] is related to the injection of an anticlockwise mover ($\downarrow$-electron) at lead $L_1$. Such an electron can not exit via lead $L_2$ unless it undergoes a spin-flip scattering. The leading process which has a finite probability amplitude to exit at $L_2$ corresponds to a situation where the injected $\downarrow$-electron first traverses a full circle starting it journey at $L_1$ via the lower arm of the ring and then crossing passed $L_2$ and reaching $L_1$ again. And then it undergoes a spin-flip scattering to bounce back as a clockwise mover and travel through the upper arm back to $L_2$ and exits the ring. Rest of the subsequent paths corresponds to the electron undergoing multiple rounds of circulation along the ring before exiting such that the total number of spin flip scattering at $L_1$ is odd and hence it is to be in ``up" state while exiting the ring via lead $L_2$.
The total transmission amplitude, which can be thought of as the sum of amplitudes of type -(i) and -(ii) discussed above is given by 
\begin{equation}
 t = \zeta[16 e^{i\phi'} \tilde{\Gamma}_1\tilde{\Gamma}_2 \Upsilon_{\uparrow 1,\Uparrow 1}(4+\tilde{\Gamma}_1^{2})(e^{i \phi_D}-1)],
 \label{tottransmit}
\end{equation}
where, 
\begin{align}
 \zeta^{-1} &= (4+\tilde{\Gamma}_1^{2})[16(e^{i\phi_D}-1)^2+4(1-e^{2i\phi_D})(\tilde{\Gamma}_1^{2}+\tilde{\Gamma}_2^{2}) \nonumber \\
 &+\tilde{\Gamma}_1^{2}\tilde{\Gamma}_2^{2}(e^{2i\phi_D}-2 \Upsilon_1 e^{i\phi_D}+1)],
 \label{zetaa}
\end{align}
and $\Upsilon_1=2|\Upsilon_{\Uparrow 1,\uparrow 1}|^2-1$ (the overlap $\Upsilon_{\Uparrow 1,\uparrow 1}$ corresponds to lead $L_1$ being attached to the region with SO field ${\bf a}_1$.\\
\textbf{Zero-pole analysis} - \textit{appearance of antiresonances} :\\
As can be seen from the expression for the transmission probability amplitude in Eq.~\ref{tottransmit}, the case for one local parallel lead is distinct from the case of both leads being local parallel in its analytic form. Eq.~\ref{tottransmit} is carrying a term $(e^{i \phi_D}-1)$ which represents first-order zeroes at $E=\phi_D=2n\pi$ resulting in Fano-type antiresonances at those points [see Fig.~\ref{fig:figg7} (b)]~\cite{fano1961effects}. These antiresonances are attributed to the interference between the two types of paths shown in Fig.~\ref{fig:figg7} (a) and hence, are directly connected to deviation of $L_1$ from its local parallel condition. Also it is interesting to note that the transmission zeros are always placed symmetrically between two maxima (around $E=2n\pi$). The pattern is related to the relative positions of the zeros and the poles of $t$ in Eq.~\ref{tottransmit}. 

It is straightforward to verify that the poles of $t$ obtained from Eq.~\ref{zetaa} are given by \begin{equation}
 \phi_D = 2n\pi -i{\rm ln} \mathcal{R}~~{\rm where}~~ \mathcal{R}=\frac{16+\Upsilon_1\tilde{\Gamma}_1^2\tilde{\Gamma}_2^2\pm\sqrt{\Delta}}{(4-\tilde{\Gamma}_1^2)(4-\tilde{\Gamma}_2^2)},
\label{poles2}
\end{equation}
where $n$ is an integer and  $\Delta=32\Upsilon_1\tilde{\Gamma}_1^2\tilde{\Gamma}_2^2+16(\tilde{\Gamma}_1^4+\tilde{\Gamma}_2^4) + (\Upsilon_1^2-1)\tilde{\Gamma}_1^4\tilde{\Gamma}_2^4$. The quantity $\mathcal{R}$ in Eq.~\ref{poles2} turns out to be real and positive in the weak tunneling limit: $\tilde{\Gamma}_{1,2}<2$. We note that the real part of the positions of zeros and poles in the complex $\phi_D$ plane are same. Hence, in the absence of the zeros, $\vert t \vert^2$ would have maxima at $E=\phi_D=2n\pi$ but due to the presence of the zeros exactly at the same positions, the original maxima split into two new symmetrically placed maxima about the transmission zeros as shown in Fig.~\ref{fig:figg7} (b).   

From Eq.~\ref{poles2}, it is evident that the locations of the poles in the complex $\phi_D$ plane have nontrivial dependence on the parameter $\Upsilon_1$ which quantifies the deviation of the polarized lead $L_1$ from its local parallel condition (i.e. when $\Upsilon_1=1$). This parameter can thought of as a control parameter which decides the width of the antiresonance. As we bring back $L_1$ to local parallel, one of the values of $\mathcal{R}$ in Eq.~\ref{poles2} approaches 1, and the imaginary component of the corresponding pole vanishes, thus, exactly cancelling the zero of $t$ in Eq.~\ref{tottransmit}. Furthermore, the complex pole, left after cancellation, coincides with the pole in $t$ for the local parallel case as expected.

In presence of a finite SB phase picked up by the electrons, the transmission probability $T$ between the leads $L_1$ and $L_2$ gets modified by $\phi_D\rightarrow \phi_D+\phi_{\rm SB}$ and so, a shift of the interference pattern [see Fig.~\ref{fig:figg7} (b)] would render a direct evidence of the presence of SB phase as noted previously.
\begin{figure}
\centering
\includegraphics[width=1.0\columnwidth]{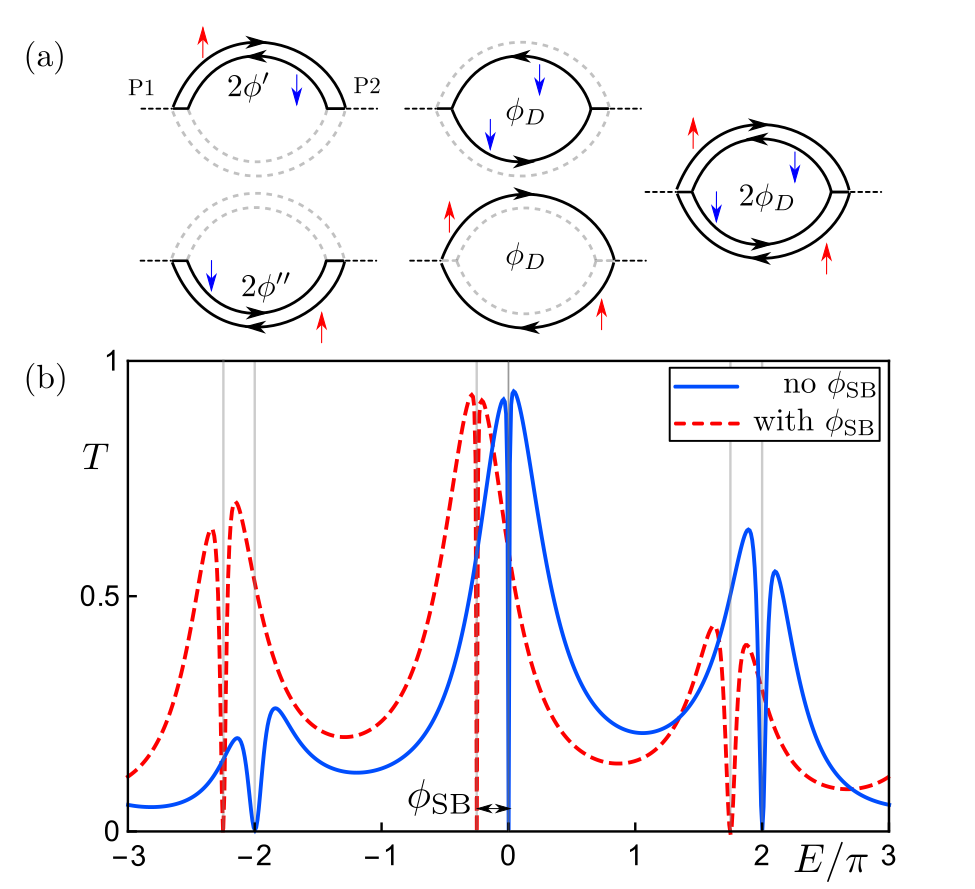}
\caption{{\bf Third scenario:} (a) The elementary closed-loop processes that, along with their multiple occurrences, contribute to the total transmission probability $T$ in a set-up with the polarization direction of both the leads deviating from the local parallel configuration. (b) The resultant interference pattern in absence (solid) and presence (dashed) of the SB phase $\phi_{\rm SB}$ [$\phi_{\rm SB}=\pi/4$ in the plot with $\hat{\bf a}_1=(0,0,1)$, $\hat{\bf a}_2=(1,0,0)$, and $\hat{\bf a}_3=(0,1,0)$] with $\phi'/\phi_D=1/3$. Lead $L_1$ is tilted by an angle of $\pi/3$ from $\hat{\bf a}_1$ keeping the azimuthal angle same while lead $L_2$ is tilted by an angle of $\pi/4$ from $\hat{\bf a}_2$ keeping the azimuthal angle same. The antiresonance points are spaced with a periodicity of $2\pi$ while the envelope of the interference pattern repeats with a periodicity of $6\pi$ as explained in the main text. The tunneling strengths are taken as $\tilde{\Gamma}_1=\tilde{\Gamma}_2=1$.}
\label{fig:figg8}
\end{figure}

\subsubsection{Third scenario : complete deviation (two leads) from local parallel condition : distorted interference pattern}

In a realistic situation, one would expect the polarization direction of both the leads to deviate from the local parallel condition when the spatial profile of the direction of the SO field on the edge is completely unknown. Following the same procedure as for the previous cases, the transmission amplitude $t$ is evaluated to be 
\begin{equation}
 t=\zeta [16\tilde{\Gamma}_1\tilde{\Gamma}_1(e^{i\phi_D}-1)(e^{i\phi'}\Upsilon_{\Uparrow 1,\uparrow 1}\Upsilon_{\Uparrow 2,\uparrow 2}+e^{i\phi{''}}\Upsilon_{\Uparrow 1,\downarrow 1}\Upsilon_{\Uparrow 2,\downarrow 2})],
\label{tottransmitpol}
\end{equation}
where
\begin{align}
 \zeta^{-1} &= (4+\tilde{\Gamma}^2_1)(4+\tilde{\Gamma}^2_2)-4\tilde{\Gamma}^2_1\tilde{\Gamma}^2_2 (\Upsilon^{\prime} e^{2i\phi'}+\Upsilon^{\prime\prime} e^{2i\phi{''}}) \nonumber \\
 &-2(16+\Upsilon_1 \Upsilon_2\tilde{\Gamma}^2_1\tilde{\Gamma}^2_2)e^{i\phi_D}+(4-\tilde{\Gamma}^2_1)(4-\tilde{\Gamma}^2_2)e^{2i\phi_D},
\label{zetatottrarnspol}
\end{align}
$\Upsilon_1=2|\Upsilon_{\Uparrow 1,\uparrow 1}|^2-1$ and $\Upsilon_2=2|\Upsilon_{\Uparrow 2,\uparrow 2}|^2-1$. $\Upsilon^{\prime} = \Upsilon_{\Uparrow 1,\downarrow 1} \Upsilon_{\downarrow 1,\downarrow 2} \Upsilon_{\downarrow 2,\Uparrow 2} \Upsilon_{\Uparrow 2, \uparrow 2} \Upsilon_{\uparrow 2, \uparrow 1} \Upsilon_{\uparrow 1, \Uparrow 1}$ and $\Upsilon^{\prime\prime} = \Upsilon_{\Uparrow 1,\uparrow 1} \Upsilon_{\uparrow 1,\uparrow 3} \Upsilon_{\uparrow 3,\uparrow 2} \Upsilon_{\uparrow 2,\Uparrow 2} \Upsilon_{\Uparrow 2, \downarrow 2} \Upsilon_{\downarrow 2, \downarrow 3} \Upsilon_{\downarrow 3, \downarrow 2} \Upsilon_{\downarrow 2, \Uparrow 1}$ with $\Upsilon_{\alpha i,\beta j}=\langle n_{\alpha i}|n_{\beta j}\rangle$ being the spinor overlap between different spinors on the HES where $\alpha,\beta=\uparrow,\downarrow$ and $i,j=1,2~\text{and}~3$; the overlap $\Upsilon_{\alpha j,\Uparrow I}$ (and its conjugate $\Upsilon_{\Uparrow I,\alpha j}$) where $I=1,2$ is defined below Eq.~\ref{htun} (definitions of $\phi'$ and $\phi''$ are given before). The quantity $\Upsilon^{\prime}$ and $\Upsilon^{\prime\prime}$ geometrically represent cyclic projections which, on the Block sphere, can be identified as hexagonal and octagonal Pancharatnam loops respectively. \\
\textbf{Position dependency of the leads} - \textit{distorted transmission pattern} :
Evidently, the phases $\phi'$ and $\phi{''}$ are dependent on the lead positions on the ring unlike their sum $\phi_D$. {\color{black}{If the arm lengths of the interferometer are equal, a distorted transmission pattern can result only due to an SB phase which is evident from Eq.~\ref{tottransmitpol} and \ref{zetatottrarnspol}.}} 
For an arbitrary value of $\phi'/\phi_D$, the poles of $t$ can have real components other than $2n\pi$, but the zeroes being pinned at $2n\pi$ (since it depends on $\phi_D$ only) results in asymmetric maxima around the antiresonance points as shown in Fig.~\ref{fig:figg8} (b). This is in distinction to the second scenario where the two maxima around each antiresonance point were symmetric [Fig.~\ref{fig:figg7} (b)] because of the coincidence of the zeroes and the real components of the poles at $2n\pi$ (see Eq.~\ref{poles2}).\\ 
\textbf{Different periodicities found in the process} - \textit{calculation of a net periodicity of the envelope of the distorted transmission pattern} :
The phases appearing in the individual terms in Eq.~\ref{zetatottrarnspol} are representatives of different closed loops formed during the spin transport from $L_1$ to $L_2$ on the interferometer [depicted in Fig.~\ref{fig:figg8} (a)] whose multiple occurrences contribute to the total transmission probability $T$ ($=|t|^2$). These phases have their own periodicity that could be different from each other depending on the lead positions, which determines the overall periodicity of the envelope of $T$ when plotted as a function of $E$. For instance, if we place our leads $L_1$ and $L_2$ such that $\phi'$ is a rational fraction of $\phi_D$ i.e. $\phi'/\phi_D=p/q$ where $p,q$ are coprime with $p<q$, the antiresonances appear in a period of $2\pi$ on the $E$ axis [see Fig.~\ref{fig:figg8} (b)] because of the zeroes of $t$ in Eq.~\ref{tottransmitpol} which bears a factor $(e^{i\phi_D}-1)$, however, the overall interference pattern is periodic with a period of $2q\pi$ if $q$ is odd and $q\pi$ if $q$ is even. In Fig.~\ref{fig:figg8} (b), we have shown the case of $p/q=1/3$ rendering a periodicity of $6\pi$ to the interference pattern when plotted against $E$.           

In presence of the SB phase, all the phase factors in the expression of $t$ (Eq.~\ref{tottransmitpol} and \ref{zetatottrarnspol}) are modified in a nontrivial way, but the total dynamical phase goes like $\phi_D\rightarrow\phi_D+\phi_{\rm SB}$ as before. This is crucial for identifying the antiresonance points which are shifted from $2n\pi$ to $2n\pi-\phi_{\rm SB}$. But note that the entire interference pattern does not experience the same overall shift unlike previously. In fact, the pattern is further distorted due to the phase factors coming from $\Upsilon'$ and $\Upsilon''$ in Eq.~\ref{zetatottrarnspol} that depend on the spin polarizations of the leads $L_1$ and $L_2$, however, the shift of the antiresonance points would still be a concrete evidence of the presence of SB phase in the system.    

\begin{figure}
\centering
\includegraphics[width=1.0\columnwidth]{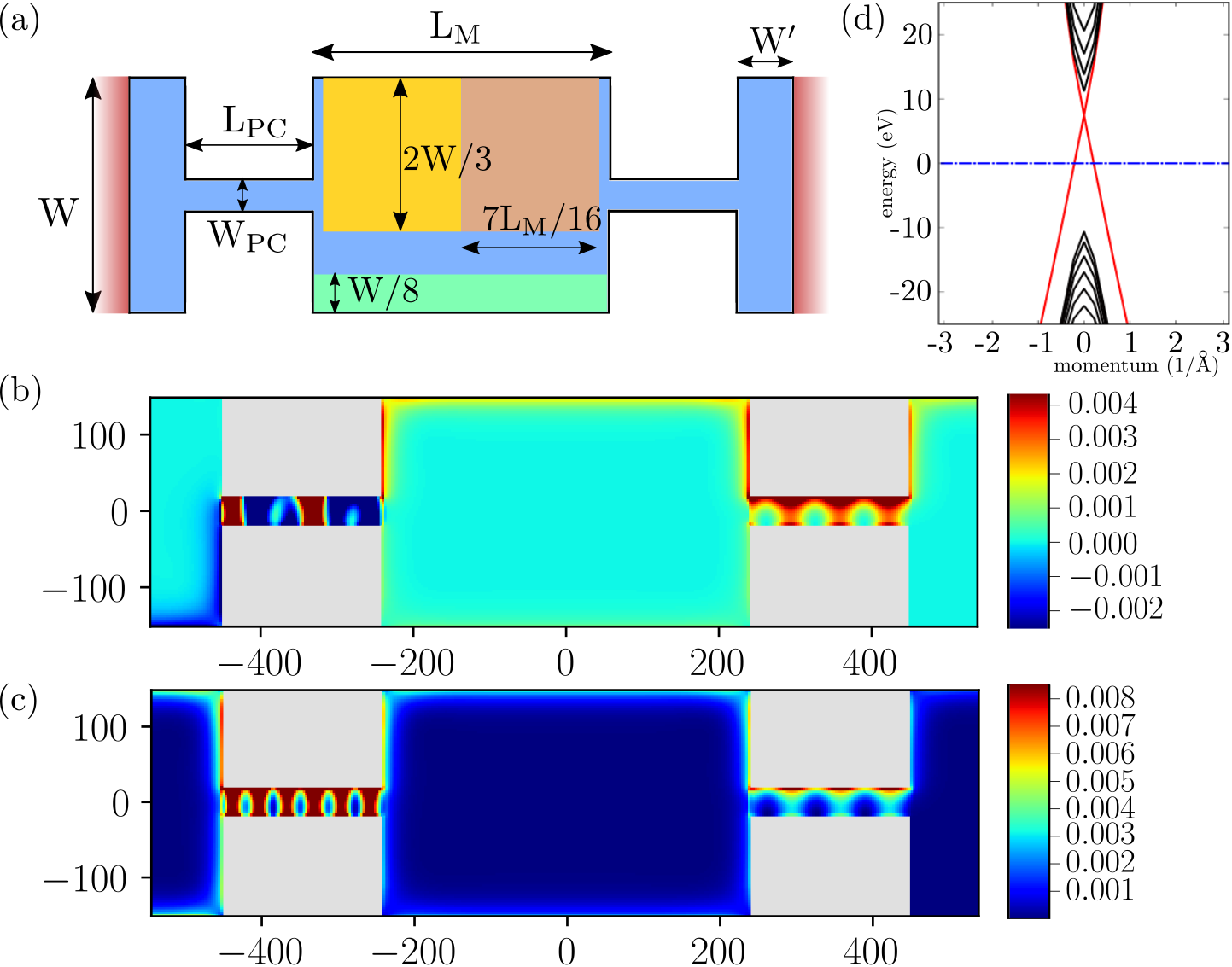}
\caption{{\color{black}{
(a) Schematic of the lattice used in KWANT. The blue region represents the system while the red the leads. The parameters used are ${\rm L}_{\rm M}=160a$, ${\rm W}=100a$, ${\rm L}_{\rm PC}=70a$, ${\rm W}_{\rm PC}=12a$, ${\rm W}'=20a$ with $a=3$nm being the lattice constant. The yellow and the brown region denote the SO region with the SO field $\hat{\bf a}=(1,0,0)$ and $(0,1,0)$ respectively. The gate voltage $V_g$ is applied (green region) across a length ${\rm L}_{\rm M}$ along the bottom edge. Spin filters (details mentioned in the text) are applied on the bottlenecks/QPCs. (b) Plot of spin densities signifying a spin-polarized injection at the left QPC (details in the text) at incident energy $E=0$. (c) Plot of the charge density showing the flow of electrons along the global edge only. No additional edges are observed to form at the boundaries of the SO regions. (d) Low-lying energy bands of the lattice model (detailed in the text) for a semiperiodic boundary condition with the red lines representing edge states with a Dirac spectrum.}}}
\label{fig:figg9}
\end{figure}

\begin{figure*}
\centering
\includegraphics[width=2.06\columnwidth]{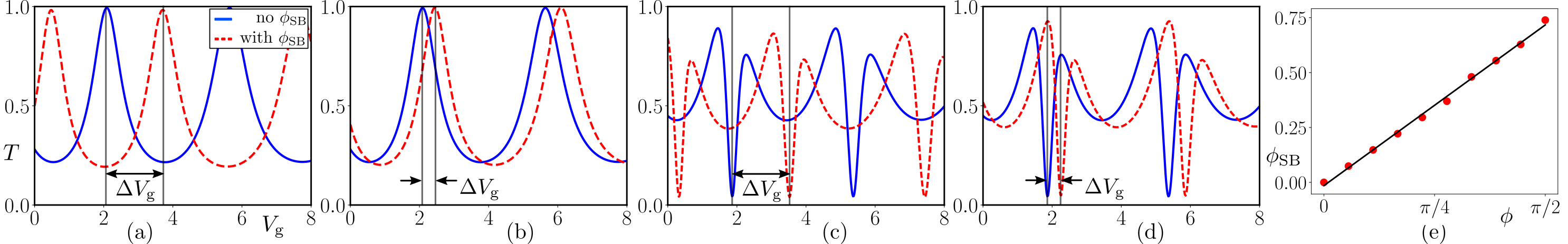}
\caption{{\color{black}{Shift of the resonance pattern for $\phi_{\rm SB}=\pi$ ($\Delta V_g=1.6$). in (a) $\phi_{\rm SB}=\pi/4$ ($\Delta V_g=0.4$) in (b). Shift of the antiresonance pattern for $\phi_{\rm SB}=\pi$ ($\Delta V_g=1.6$) in (c) and $\phi_{\rm SB}=\pi/4$ ($\Delta V_g=0.4$) in (d). (e) Variation of the SB phase as a function of the relative orientation of the SO field (parameterized by the relative angle $\phi$ mentioned in the text) in the consecutive barriers [shown in Fig.~\ref{fig:figg9} (a)] showing the expected linear behavior $\phi_{\rm SB}=\phi/2$. The plots for resonances are achieved by setting $\beta=0.5$ on the left bottleneck while those for the antiresonances are obtained by setting $\beta=0.8$ on the left bottleneck. Both cases have $\beta=0$ on the right bottleneck and share the same properties as noted in Fig.~\ref{fig:figg9}.}}}
\label{fig:figg10}
\end{figure*}


{\color{black}{\section{Numerical analysis using KWANT}\label{secfour}

To demonstrate the phenomenon of $S_z$ nonconservation in a realistic interferometer set-up following our prescription, we simulate a lattice model and study its transport properties using the KWANT package~\cite{groth2014kwant}. This is a minimal set-up to capture the essential physics where we have engaged a modified version of the BHZ model that will be discussed shortly. It is to be noted that in our case, the bulk quantum spin Hall state is spread over the interferometer region whose bulk $S_z$ symmetry is broken (but time-reversal symmetry is kept) endowing the interferometer region with a possibility of fragmenting into multiple patches each having a distinct spin quantization axis and thus, resulting in the edge states formed on the interferometer arms to acquire a nontrivial SB phase. As all distinct $S_z$ symmetry breaking bulk states are degenerate, such regions of $S_z$ symmetry broken bulk states could appear spontaneously in the system. We further note no additional edge states are formed at the interfaces between these distinct patches and only a dominant single edge appears at the global boundary of the full region which is evident from the charge and spin density plots presented in Fig~\ref{fig:figg9}.

In what follows, we will first discuss the geometry of the lattice and then provide details of the model considered. The geometry  is similar to that of Ref.~\onlinecite{SBP1} and shown in Fig~\ref{fig:figg9} (a) with the parameters mentioned therein. The boundary of the lattice is specified with coordinates $(x,y)$ where $y \in [-Y(x),Y(x)]$ and
\begin{align}
Y(x) = \begin{cases}
  W/2, & \text{if } x < -(L_{\rm M}/2 + L_{\rm PC}), \\
  W_{\rm PC}/2, & \text{if}-(L_{\rm M}/2 + L_{\rm PC}) \leq x \leq -L_{\rm M}/2, \\
  W/2  & \text{if} -L_{\rm M}/2 < x < L_{\rm M}/2, \\
  W_{\rm PC}/2  & \text{if~} L_{\rm M}/2 \leq x \leq (L_{\rm M}/2 + L_{\rm PC}), \\
  W/2  & \text{if~} x > (L_{\rm M}/2 + L_{\rm PC}). 
\end{cases}
\end{align} 
An appropriate setting of the dimensions of the bottlenecks facilitates the desired backscattering enabling these regions to serve as QPCs (P1 and P2 as shown in Fig.~\ref{fig:figg5}) to which extended leads are connected for current measurements.  

Let us now turn to the description of our model that is simulated using KWANT.
The original BHZ model is specified by the Hamiltonian  
\begin{align}
H_{\rm BHZ} = -D k^2 + A k_x \sigma_z \tilde{\sigma_x} - A k_y \tilde{\sigma_y} + (M-B k^2) \tilde{\sigma_z},
 \label{BHZham}
\end{align}
where $\sigma$ and $\tilde{\sigma}$ denote the Pauli matrices to describe the spins (up or down along $S_z$) and the orbitals (electron type or hole type) respectively and $A$, $B$, $D$ and $M$ are material dependent parameters. 
{\color{black}{As noted previously, distinct regions of $S_z$ symmetry broken bulk states could appear spontaneously in the system. This is evident from the structure of the model Hamiltonian for the bulk of the quantum spin Hall system given in Eq.~\ref{BHZham} and the fact that a replacement of $\sigma_z$ in the second term of Eq.~\ref{BHZham} by ${\hat{\bf n}}.{\bf \sigma}$, where ${\hat{\bf n}}$ is a unit vector pointing in an arbitrary direction in the spin space, does not alter the spectrum  of the BHZ Hamiltonian or the presence of the helical edge states. This leads us to consider the bulk Hamiltonian of the form 
\begin{align}
\tilde{H}_{\rm BHZ} = -D k^2 + A k_x ({\hat{\bf n}}.{\bf \sigma}) \tilde{\sigma_x} - A k_y \tilde{\sigma_y} + (M-B k^2) \tilde{\sigma_z},
 \label{BHZham2}
\end{align}
and $\hat{\bf n}=\hat{\bf a}_i$ in a region specified by the SO vector $\hat{\bf a}_i$.}}

In order to obtain a neat interference pattern as a function of the gate voltage $V_g$, applied to the bottom edge of the lattice in a region of width $W/8$ extending from $x=-L_M/2$ to $x=L_M/2$ [see Fig.~\ref{fig:figg9} (a)], the incident energy of the electrons is adjusted to $E=0$, neither close to the Dirac point nor the band edges. Edge states are observed in the low-lying spectrum of the model calculated with semiperiodic boundary condition and displayed in Fig.~\ref{fig:figg9} (d).

The tight-binding version of the BHZ Hamiltonian (Eq.~\ref{BHZham2}) that applies to a region with an SO field given by $\hat{\bf n}=\hat{\bf a}_i$ on a square lattice is constructed with a basis of two sites (representing the orbitals) using $k^2 = 2 a^{-2}[2-\cos(k_x a)-\cos(k_y a)]$, $k_x = a^{-1} \sin(k_x a)$, and $k_y = a^{-1} \sin(k_y a)$ which reads
\begin{equation}
H_{\rm tb} = \sum_i (c_i^{\dagger} H_{i,i+a_x} c_{i+a_x} + c_i^{\dagger} H_{i,i+a_y} c_{i+a_y} + {\rm h.c.}) + c_i^{\dagger} H_{ii} c_i,
\end{equation}
where $c_i^{\dagger} \equiv (c^{\dagger}_{i,s,\uparrow},c^{\dagger}_{i,p,\uparrow},c^{\dagger}_{i,s,\downarrow},c^{\dagger}_{i,p,\downarrow})$ denotes the set of creation operators for the electrons in $s$ and $p$ orbital with $\uparrow$ and $\downarrow$ spins at site $i$ with coordinates $i=(i_x,i_y)$; $a_x = a(1,0)$ and $a_y = a(0,1)$ are the lattice vectors with $a$ being the lattice constant. Each of the terms, $H_{ii}$ and $H_{i,i+a_x(a_y)}$, is a $4 \times 4$ block matrices defined by
\begin{align}
H_{ii} &= -\frac{4 D}{a^2} - \frac{4 B}{a^2} \tilde{\sigma_z} + M \tilde{\sigma_z} \nonumber \\
H_{i,i+a_x} &= \frac{D + B \tilde{\sigma_z}}{a^2} + \frac{A ({\hat{\bf a}}_i.{\bf \sigma}) \tilde{\sigma_x}}{2 i a} \nonumber \\
H_{i,i+a_y} &= \frac{D + B \tilde{\sigma_z}}{a^2} + \frac{i A \tilde{\sigma_y}}{2 a}.
\label{bhzlattice}
\end{align}
We have used the standard parameters for the HgTe/CdTe quantum wells which are $A = v_F = 364.5$ nm meV, $B = -686$ ${\rm nm}^2$ meV, $D = -512$ ${\rm nm}^2$ meV, and $M = -10$ meV~\cite{Konig766}. The lattice constant is set to $a =3$ nm to obtain a reasonable band structure~\cite{SBP1}. 
As argued before, a finite value of the SB phase requires three distinct configurations of the SO field, one of which is already accommodated in the BHZ model with orientation $\hat{\bf a}_1=(0,0,1)$. The two additional SO regions have the SO field oriented along $\hat{\bf a}_2=(1,0,0)$ and $\hat{\bf a}_3(\phi)=(\cos\phi,\sin\phi,0)$ and placed in a sequence [each region having a length of $7L_{\rm M}/16$ and a width of $2W/3$ as shown in Fig.~\ref{fig:figg9} (a)]. 
 
The leads are constructed with the same Hamiltonian as in Eq.~\ref{BHZham}. However, to observe the desired antiresonances in the interference pattern, spin-polarized electrons are to be injected into the system (resonances can also be achieved with spin-polarized leads provided they are kept in local parallel condition as explained previously). This is implemented by deploying spin filters on the bottlenecks modeled by the lattice Hamiltonian 
\begin{align}
 H_{ii} &= -\frac{4 D}{a^2} - \frac{4 B}{a^2} \tilde{\sigma_z} + M \tilde{\sigma_z} \nonumber \\
 H_{i,i+a_x} &= \frac{D + B \tilde{\sigma_z}}{a^2} + \frac{A {\tau}_a(\beta) \tilde{\sigma_x}}{2 i a} \nonumber \\
 H_{i,i+a_y} &= \frac{D + B \tilde{\sigma_z}}{a^2} + \frac{i A \tilde{\sigma_y}}{2 a},
 \label{latticelead}
\end{align}
where ${\tau}_a(\beta)=(1+\beta\sigma_a)/2$ [$a\in \{x,y,z\}$] decides the spin polarization of the injected electrons. The filter selectively injects/receives spins aligned in the $\pm S_a$ direction depending on the value of the parameter $\beta\in[-1,1]$. We set $\beta=0.5$ for the filter placed on the left bottleneck and since to capture the essential features of our interferometer (which are the resonances and antiresonances), it is sufficient to place the filter at the left bottleneck only, we accordingly set $\beta=0$ for the right bottleneck. Plots of the charge and spin densities over the entire system as shown in Fig.~\ref{fig:figg9} (b)-(c) reveal the behavior of the bottleneck regions as QPC. As $\beta>0$ in the left QPC, strong backscattering in spin $\downarrow$ channel results as evident in Fig.~\ref{fig:figg9} (b) which also shows a dominant $\uparrow$ transmission via the other QPC (note that the density of $\uparrow$ electrons is slightly suppressed in the lower edge of the interferometer because of the backscattering in the right QPC). Fig.~\ref{fig:figg9} (c) shows the flow of the spin-polarized electrons along the edges. Resonances are observed when the leads are local parallel which amounts to setting ${\tau}_a={\tau}_z$ in Eq.~\ref{latticelead}. Antiresonances result from deviations from the local parallel configuration which is achieved by setting $\tau_a=\tau_x$.  

Equipped with this interferometer set-up, we now demonstrate the phenomenon of $S_z$-non conservation on the QSH edges by measuring the conductance in presence of an appropriate environment of the SO fields as mentioned previously. The corresponding results are summarized in Fig.~\ref{fig:figg10}. 
A plot of the SB phase as a function of $\phi$ is presented in Fig.~\ref{fig:figg10} (e) which reflects the desired relation $\phi_{\rm SB}=\phi/2$. The plots in Fig.~\ref{fig:figg10} (b) and (d) correspond to the SB phase of $\phi_{\rm SB}=\pi/4$ i.e. when $\hat{\bf a}_3=(0,1,0)$.
Similarly, a topological SB phase of can be achieved by having all the SO field directions restricted to lie on the $y-z$ plane in the Bloch sphere, however, as stressed before, the details of the variation matter. To illustrate this, we consider, besides $\hat{\bf a}_1=(0,0,1)$, four other distinct SO barriers on the lattice given by $\hat{\bf a}_{2,3,4,5}$ where $\hat{\bf a}_j=(0,\sin\xi_j,\cos\xi_j)$ with $\xi_2=\pi/3, \xi_3=3\pi/4, \xi_4=5\pi/4$ and, $\xi_5=5\pi/3$. This configuration amounts to $\phi_{\rm SB}=\pi$ as observed in Fig.~\ref{fig:figg10} (a) and (c).
}}


\section{Discussion and Conclusion}\label{secfive}

Studies on edge transport in quantum spin Hall systems have primarily considered situations where the $S_z$ component of the spin is conserved. However, it is only the time reversal symmetry which is required to protect the edge state. The present paper explores the consequences of relaxing the $S_z$ conservation by considering a generic profile of the spin-orbit (SO) field along a pristine edge. As a result, we observe spin Berry (SB) phase accumulated by the electrons flowing along the edge, a finite value of which warrants $S_z$ nonconservation and has notable effects on edge transport.

To measure such a phase, it is essential to employ an interferometric set-up for which we consider a ring geometry of the edge state tunnel-coupled to two spin-polarized leads that serves as a two-path interferometer. {\color{black}{In a realistic experimental set-up, realizing a ring geometry of the edge state will involve a double-QPC geometry as shown in (Fig.~\ref{fig:figg5}). In a recent experiment, a single QPC in a quantum spin Hall edge has been devised~\cite{strunz2019interacting} and hence, our proposed set-up is also realizable in similar types of experiments.}} Motivated by Pancharatnam's construct of geometric phase, we present the minimal criteria for the SO profile to lead to a finite accumulation of SB phase. Furthermore, we provide an explicit derivation of the expression for the SB phase in terms of the SO field configurations using a transfer matrix approach that constitutes one of the main results of the paper. The compact form of the transfer matrix presented in this paper is instructive in developing an understanding of the geometric aspect of the spin dynamics of itinerant electrons.  

In our set-up, introduction of the spin-polarized leads results in sharp antiresonance in the transmission probability which, in presence of a finite SB phase, get shifted by an amount equal to the SB phase. We analyze three distinct situations depending on various possible orientations of the polarization directions of the leads that leave pronounced effects on the overall pattern of the transmission probability including its periodicity, however, the features of antiresonance remain in all such cases.

{\color{black}{As a final remark, we note that to measure the shift due to the SB phase, a reference point needs to be identified for which the dynamical phase $\phi_{\rm D}=Lk$ can be used as a marker ($L$ is the total length of the interferometer (sum of the upper and lower arm) and $k$ is the wavevector of the incident electron) as follows. When a small bias voltage $V_g$ is applied, it corresponds to an wavevector $k={eV_g}/({\hbar v_F})$ where $eV_g$ is the energy of the incident electron with respect to the Fermi energy of the edge state. In presence of an SB phase, the total interferometric phase is given by $\phi=\phi_{\rm D} + \phi_{\rm SB}$. If we measure the differential conductance (or equivalently, the total transmission probability $T$) as a function of the bias voltage $V_g$ such that the interference pattern repeats itself over a voltage difference of $\delta V_g=(V_1-V_2)$, it amounts to setting the corresponding phase difference $\delta\phi=2\pi$ where $\delta\phi = L(k_1-k_2) = {Le}(V_1-V_2)/({\hbar v_F}) \equiv {Le\delta V_g}/({\hbar v_F})$. This provides an estimate of $v_F$ that can be further used to compute the value of the wavevector $k$. A direct measurement of $\phi_{\rm SB}$ then immediately follows since the rest of the parameters are known ($L$ can be estimated from the scanning electron microscope image of the device). In absence of the SB phase, $Lk=LeV_g/(\hbar v_F)=2\pi$ provides the expected position of the first resonance in terms of the bias voltage $V_g$, and any excess shift from this position captured in a scan over $V_g$ would be solely attributed to the SB phase.}}            


\section{Acknowledgments} 

The authors thank Poonam Mehta for illuminating discussions and scrutinizing the manuscript. V.A. acknowledges financial support from University Grants Commission, India and K.R. acknowledges sponsorship, in part, by the Swedish Research Council. S.D. would like to acknowledge the MATRICS grant (Grant No. MTR/2019/001043) from 
Science and Engineering Research Board, India for funding.


\appendix

\section{\\Derivation of SB phase from the product of transfer matrices}\label{appA}

Here we will present the derivation of Eq.~\ref{tri} given in the main text. In particular, we analyze the terms in Eq.~\ref{sangle} that lead to the simplified expression of Eq.~\ref{tri}.

We start with the scalar triple product in the last term of Eq.~\ref{sangle} and write it as 
\begin{align}
 [{\hat{\bf D}}_{13},{\hat{\bf D}}_{32},{\hat{\bf D}}_{21}] 
 = \frac{\mathcal{B}}{\sin2\theta_{21}\sin2\theta_{13}\sin2\theta_{32}},
\end{align}
where
\begin{align}
\mathcal{B} &= |\hat{{\bf a}}_1\cdot(\hat{{\bf a}}_2 \times \hat{{\bf a}}_3)|^2 \nonumber \\ 
 &= \Big[1-({\hat{\bf a}}_1\cdot{\hat{\bf a}}_2)^2-({\hat{\bf a}}_2\cdot{\hat{\bf a}}_3)^2
 -({\hat{\bf a}}_3\cdot{\hat{\bf a}}_1)^2 \nonumber \\
 &+ 2({\hat{\bf a}}_1\cdot{\hat{\bf a}}_2)({\hat{\bf a}}_2\cdot{\hat{\bf a}}_3)({\hat{\bf a}}_3\cdot{\hat{\bf a}}_1)\Big].
\end{align}
This expression is obtained using ${\hat{\bf D}}_{ij} = ({\bf a}_i\times{\bf a}_j)/|{\bf a}_i\times{\bf a}_j|$. The last term of Eq.~\ref{sangle} then becomes
\begin{align}
 [{\hat{\bf D}}_{13},{\hat{\bf D}}_{32},{\hat{\bf D}}_{21}]&\sin\theta_{13}\sin\theta_{32}\sin\theta_{21} \nonumber \\
 &= \frac{\mathcal{B}}{\Big(8\cos\theta_{13}\cos\theta_{32}\cos\theta_{21}\Big)} \nonumber \\
 &\equiv \frac{1-x^2-y^2-z^2+2xyz}{8\cos\theta_{13}\cos\theta_{32}\cos\theta_{21}},
\label{lasteterm1}
\end{align}
where
\begin{align} 
& x\equiv {\hat{\bf a}}_2\cdot{\hat{\bf a}}_1=\cos2\theta_{21}, y\equiv {\hat{\bf a}}_3\cdot{\hat{\bf a}}_2=\cos2\theta_{32}, \nonumber \\
& z\equiv {\hat{\bf a}}_1\cdot{\hat{\bf a}}_3=\cos2\theta_{13}.
\label{xyz} 
\end{align}
Similarly, the first term of Eq.~\ref{sangle} can be written as
\begin{align}
 &\cos\theta_{13}\cos\theta_{32}\cos\theta_{21} \nonumber \\
& = \frac{(1+\cos2\theta_{13})(1+\cos2\theta_{32})(1+\cos2\theta_{21})}{8\cos\theta_{13}\cos\theta_{32}\cos\theta_{21}} \nonumber \\
 &\equiv\frac{1+x+y+z+xy+yz+zx+xyz}{8\cos\theta_{13}\cos\theta_{32}\cos\theta_{21}},
 \label{firstterm1}
\end{align}
and similarly, the second, third, and fourth term of Eq.~\ref{sangle} become
\begin{align}
({\hat{\bf D}}_{13}\cdot{\hat{\bf D}}_{32})&\sin\theta_{13}\sin\theta_{32}\cos\theta_{21} \nonumber \\
 &= \frac{xyz-x^2+yz-x}{8\cos\theta_{13}\cos\theta_{32}\cos\theta_{21}},
\label{secondterm1}
\end{align}
\begin{align}
 ({\hat{\bf D}}_{32}\cdot{\hat{\bf D}}_{21}) & \cos\theta_{13}\sin\theta_{32}\sin\theta_{21} \nonumber \\
 & = \frac{xyz-z^2+xy-z}{8\cos\theta_{13}\cos\theta_{32}\cos\theta_{21}},
 \label{thirdterm1}
\end{align}
and 
\begin{align}
 ({\hat{\bf D}}_{21}\cdot{\hat{\bf D}}_{13})&\sin\theta_{13}\cos\theta_{32}\sin\theta_{21} \nonumber \\
 & = \frac{xyz-y^2+zx-y}{8\cos\theta_{13}\cos\theta_{32}\cos\theta_{21}}
 \label{fourthterm1}
\end{align}
respectively. Finally, combining all these terms, Eq.~\ref{sangle} simplifies to
\begin{align}
 \cos\alpha &= \frac{1+x+y+z}{4\cos\theta_{13}\cos\theta_{32}\cos\theta_{21}} \nonumber \\
 &\equiv\frac{1+x+y+z}{\Delta}, \nonumber \\
\Rightarrow \tan\alpha &= \frac{\sqrt{\Delta^2-(1+x+y+z)^2}}{1+x+y+z}.
 \label{term1}
\end{align}
Noting
\begin{align}
 \Delta^2 &= 16\cos^2\theta_{13}\cos^2\theta_{32}\cos^2\theta_{21} \nonumber \\
 &=2(1+x)(1+y)(1+z) \nonumber \\
 &= 2(1+x+y+z+xy+yz+zx+xyz) \nonumber \\
 &= (x+y+z)^2+2(x+y+z)+1 \nonumber \\
 &+ 1 + 2xy+2yz+2zx-(x+y+z)^2 + 2xyz \nonumber \\
 &= (1+x+y+z)^2 + (1-x^2-y^2-z^2+2xyz),
\end{align}
we arrive at
\begin{align}
 \tan\alpha &= \frac{\sqrt{1-x^2-y^2-z^2+2xyz}}{1+x+y+z},
\end{align}
which, in terms of ${\hat{\bf a}}_{1,2,3}$~, is given in Eq.~\ref{tri} of the main text.


\bibliography{references}
\end{document}